\documentclass[graybox]{svmult}

\usepackage{type1cm}        
\usepackage{makeidx}         
\usepackage{graphicx}        
\usepackage{multicol}        
\usepackage[bottom]{footmisc}
\usepackage{aastex_macro}

\usepackage{newtxtext}       %
\usepackage{newtxmath}       
\usepackage{soul}

\usepackage{float}
\usepackage{graphicx}
\usepackage{subcaption}

\makeindex             

\definecolor{gri}{rgb}{0.19, 0.55, 0.91}
\definecolor{max}{rgb}{0.9, 0.0, 0.6}

\definecolor{ste}{rgb}{0., 0.26, 0.15}

\definecolor{mau}{rgb}{1, 0.5, 0.2}

\begin{document}

\title*{Anomaly detection in Astrophysics:\\
a comparison between unsupervised Deep and Machine Learning on KiDS data}
\titlerunning{Anomaly detection in Astrophysics: Deep vs Machine Learning models}

\author{Maurizio D'Addona, Giuseppe Riccio, Stefano Cavuoti, Crescenzo Tortora, Massimo Brescia}
\authorrunning{D'Addona et al. 2020}

\institute{Maurizio D'Addona \at  Department of Physics, University of Naples  Federico II, Strada Vicinale Cupa Cintia, 21, I-80126 Napoli, Italy. \email{mauritiusdadd@gmail.com} 
\and Giuseppe Riccio \at INAF - Astronomical Observatory of Capodimonte, Salita Moiariello 16, I-80131 Napoli, Italy \email{giuseppe.riccio@inaf.it} \and Stefano Cavuoti \at  Department of Physics, University of Naples  Federico II, Strada Vicinale Cupa Cintia, 21, I-80126 Napoli, Italy.  \at INAF - Astronomical Observatory of Capodimonte, Salita Moiariello 16, I-80131 Napoli, Italy \email{stefano.cavuoti@gmail.com}
\and Crescenzo Tortora \at INAF - Astronomical Observatory of Capodimonte, Salita Moiariello 16, I-80131 Napoli, Italy \at INAF - Osservatorio Astrofisico di Arcetri, Largo Enrico Fermi 5, 50125 Firenze, Italy \email{crescenzo.tortora@inaf.it}  
\and Massimo Brescia \at INAF - Astronomical Observatory of Capodimonte, Salita Moiariello 16, I-80131 Napoli, Italy \email{massimo.brescia@inaf.it}}

%
\maketitle
\textit{Preprint version of the manuscript to appear in the Volume ``Intelligent Astrophysics'' of the series ``Emergence, Complexity and Computation'', Book eds. I. Zelinka, D. Baron, M. Brescia, Springer Nature Switzerland, ISSN: 2194-7287}\newline
\abstract*{
Every field of Science is undergoing unprecedented changes in the discovery process, and Astronomy has been a main player in this transition since the beginning. The ongoing and future large and complex multi-messenger sky surveys impose a wide exploiting of robust and efficient automated methods to classify the observed structures and to detect and characterize peculiar and unexpected sources. We performed a preliminary experiment on KiDS DR4 data, by applying to the problem of anomaly detection two different unsupervised machine learning algorithms, considered as potentially promising methods to detect peculiar sources,  a Disentangled Convolutional Autoencoder and an Unsupervised Random Forest. The former method, working directly on images, is considered potentially able to identify peculiar objects like interacting galaxies and gravitational lenses. The latter instead, working on catalogue data, could identify objects with unusual values of magnitudes and colours, which in turn could indicate the presence of singularities.
}

\abstract{
Every field of Science is undergoing unprecedented changes in the discovery process, and Astronomy has been a main player in this transition since the beginning. The ongoing and future large and complex multi-messenger sky surveys impose a wide exploiting of robust and efficient automated methods to classify the observed structures and to detect and characterize peculiar and unexpected sources. We performed a preliminary experiment on KiDS DR4 data, by applying to the problem of anomaly detection two different unsupervised machine learning algorithms, considered as potentially promising methods to detect peculiar sources,  a Disentangled Convolutional Autoencoder and an Unsupervised Random Forest. The former method, working directly on images, is considered potentially able to identify peculiar objects like interacting galaxies and gravitational lenses. The latter instead, working on catalogue data, could identify objects with unusual values of magnitudes and colours, which in turn could indicate the presence of singularities.
}

\section{Introduction}
\label{sec:1}
Due to the rapid growth in volume and complexity of astronomical datasets, Machine Learning (ML) paradigms are gaining a key role within the data exploration and analysis. They are demonstrated as valid mechanisms to find hidden correlations among data and to discover rare and unexpected structures that do not fit those relations~\cite{Baron2017,brescia2018,fluke2019}. The latter, considered as outliers of a data distribution, can be of various nature and may have different degrees of scientific relevance: they can be artifacts produced by anomalies in the data processing pipelines or in the observing conditions, as well as peculiar objects underlining special and rare astronomical events, whose detection may improve the scientific knowledge of relevant physical phenomena.\\
Machine learning paradigms are mainly divided in two main classes, respectively, supervised and unsupervised methods. 
While in the supervised case, an a-priori \textit{Knowledge Base} is needed to train the algorithms, unsupervised methods can learn the complex relationships among data, without inferring any known information and with a minimum of human supervision. Therefore, it is evident that unsupervised methods are the most suitable to detect anomalies. In particular, we focus on two specific models: an unsupervised variant of random forests (\textit{Unsupervised Random Forest}, or \textit{URF})~\cite{Shi2006} and a hybrid type of autoencoder (\textit{Disentangled Convolutional Autoencoder}, or \textit{DCA}), which exploits the disentangling property of a variational autoencoder~\cite{chen2018},  but preserving the structure of a standard convolutional autoencoder~\cite{guo2017}.

In recent years both methods have successfully been used in the astrophysical context. For example, \textit{Tuccillo et al.}~\cite{Tuccillo2018} validated the former method on both analytic profiles and real galaxy images. \textit{Baron et al.}~\cite{Baron2017} used a \textit{URF} on galaxy spectra from the \textit{Sloan Digital Sky Survey} (SDSS), finding objects with extreme emission line ratios, abnormally strong absorption lines, extremely reddened galaxies and other peculiar objects. \textit{Reis et al.}~\cite{Reis2018a} applied this method to infrared spectra of stars, showing that the metric defined in this algorithm traced the physical properties of the stars. Finally, \textit{Reis et al.}~\cite{Reis2018} also discovered $31$ new redshifted broad absorption line quasars within \textit{SDSS} spectral data. Concerning the DCA model, a very similar architecture was successfully applied to radio data to disentangle noise signal contamination, revealing emissions from air showers, thus enabling accurate measurements of cosmic particle kinematics and identity~\cite{erdmann2019}. More in general, such models are faster, compared to other traditional profile fitting methods, can be easily adapted to more simple/complex models and could be used to detect peculiar substructures, such as strong gravitational lenses and galaxy mergers.\\
In this preliminary work we first use \textit{DCA} on synthetic images in order to evaluate its theorethical performance, then we apply both methods on real image cutouts and catalogue counterparts. In particular, in Section \ref{sec:3} we describe the use of a DCA to perform an outlier detection using images extracted from the 4th Data Release of the European Southern Observatory (ESO) Kilo Degree Survey (KiDS) ~\cite{Kuijken2019}. Then, for the same purpose, in Section \ref{sec:4} we describe the use of an URF on the same subset of objects, but using photometric data, always extracted from the KiDS DR4. Finally in Sections \ref{sec:5} and \ref{sec:6} we discuss the results and compare the performance of the two methods. 

\section{Data Preparation}
\label{sec:2}
In order to validate the \textit{DCA} model and assess its performance we generated three sets of $20,000$ synthetic images of $64 \times 64$ pixels, using three different models of surface brightness profile of galaxies that are further described in sections \ref{subsec:3_1}. These images have a dynamic range between 0 and 1. A Gaussian noise, drawn from a folded normal distribution with standard deviation of $\sigma_{noise} = 0.09$, has also been added to each image and the value of the standard deviation has been chosen to maintain the $\sim 99\%$ of the values within the $30\%$ of the dynamic range. The generated noise has a mean value $\mu_{noise} \approx 5\cdot {10}^{-2}$ that corresponds to the $5\%$ of the maximum value of the dynamic range of the image.\\
The real data selected to perform our tests on both methods are extracted from the KIDS Data Release 4~\cite{Kuijken2019}. In particular, we randomly extracted a subset of object cutouts from the tiles that are in common with the DR3 data release \cite{dejong2017} and using the DR4 photometry in the related catalogue.\\
For the photometry we used the Gaussian Aperture and PSF (GAaP) magnitudes in the four bands $u$, $g$, $r$, $i$, with the minimum aperture of $1.0$ arcsec and the corresponding automatic minimal aperture magnitudes $u_{auto}$, $g_{auto}$, $r_{auto}$, $i_{auto}$, which are also corrected for the galactic extinction. In addition to these features we also included all colours and magnitude ratios~ \cite{disanto2018}, derived from all the above magnitudes, resulting in a total of $36$ photometric features. From this dataset we excluded all objects with missing data in any of the photometric bands. We also applied a minimum set of magnitude cuts, in order to remove the objects lying in the tails of the distributions: $16 < i < 22$ and $16 < r < 22$. The result is a dataset of $400,000$ objects.\\
For each object a cutout of $32 \times 32$ pixels (corresponding to $\sim 6.7 \times 6.7$ arcsec) has been extracted from the corresponding photometrically and astrometrically calibrated $r$ band coadded tiles. The size of the cutouts has been chosen so that almost all of them contain only the central object, while preserving a sufficient amount of surrounding pixels and angular size. All pixel values of the cutouts have been normalised between $0$ and $1$.

About the $90\%$ of these objects was also present in the candidate quasars (QSOs) catalogue, produced for the 3rd Data Release of KiDS, containing a mixed set of stars, QSOs and galaxies, classified with Machine Learning \cite{Nakoneczny2019}. The two catalogues were cross-matched, resulting in a subset of $\sim1100$ QSOs and $\sim260$ stars with a reliable classification, considered a useful information to take into account in the evaluation of the anomaly detection experiment results.

\section{Disentangled Convolutional Autoencoders}
\label{sec:3}

\begin{figure}
\centering
\includegraphics[width=10cm]{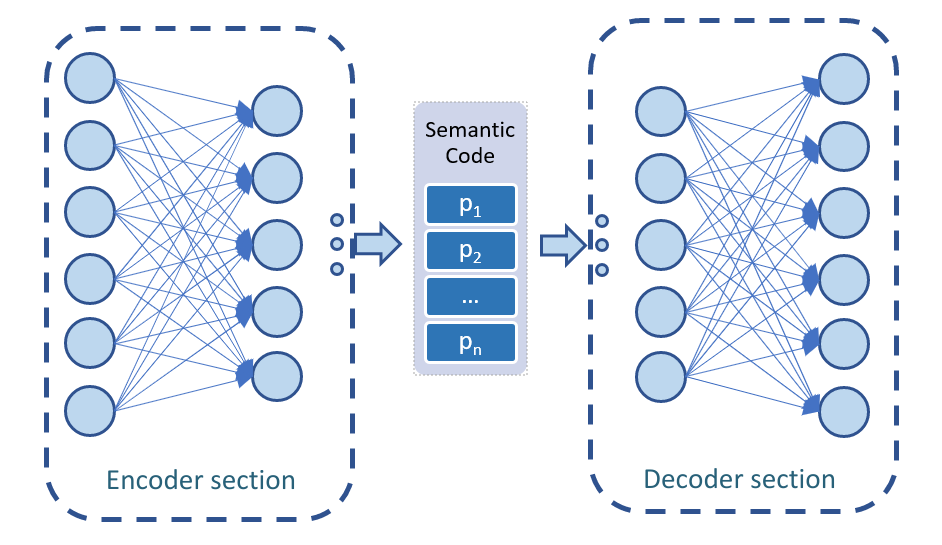}
\caption{Representation of the basic structure of an autoencoder\label{fig:autoenc_1}}
\end{figure}

Autoencoders are a particular type of neural network used to learn data codings by efficiently mapping high-dimensional inputs into low-dimensional encoded vectors and reconstructing the input data from the encoded vector only~\cite{Goodfellow-et-al-2016}. By forcing the low-dimensional representation, or \textit{latent space}, to have less dimensions than the input data, the network is forced to learn useful features from the data and, through the use of the backpropagation algorithm, in combination with a smooth loss function, the content of the latent space is iteratively adapted, in order to achieve a good reconstruction performance.
For such reasons the autoencoder is able to perform the feature extraction and dimensionality reduction tasks in a completely unsupervised fashion. The basic structure of an autoencoder consists of two sections (Fig. \ref{fig:autoenc_1}):
\begin{itemize}
\item{An \textbf{encoder} that maps the input data into \textit{semantic code vectors}, that live in a so called \textit{latent space}.}
\item{A \textbf{decoder} that learns to decompress the \textit{semantic code vectors} from the \textit{latent space} back to the input space, producing a reconstructed representation of the input.}
\end{itemize}

\begin{figure}[t]
\centering
\includegraphics[width=10cm]{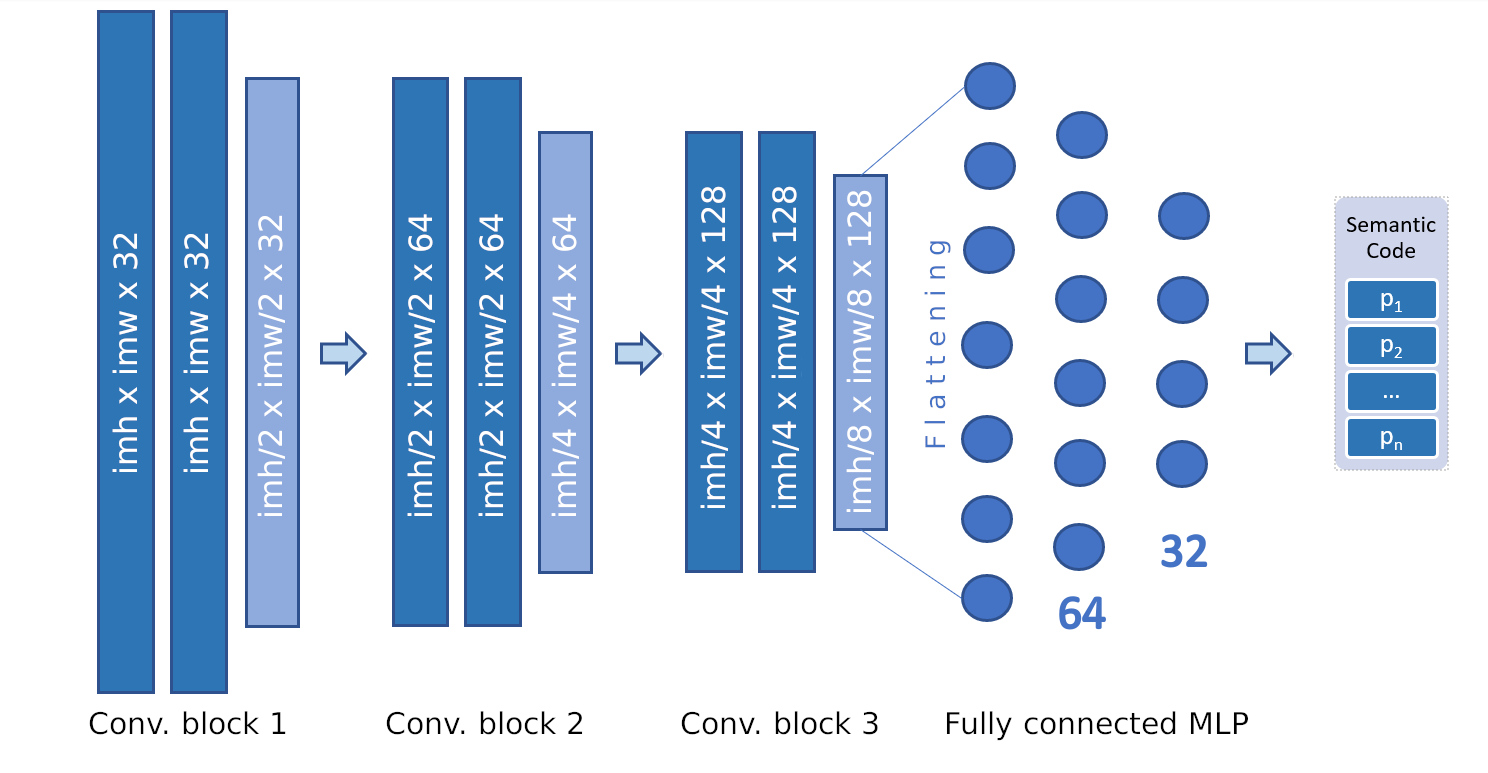}
\caption{The structure of the encoder used in our experiments: there are three convolutional blocks followed by a fully connected MLP with two hidden layers. Each block has two convolutional layers followed by a $2 \times 2$ max-pooling operation. \textit{imw} and \textit{imh} are respectively width and height of the input images.\label{fig:enc_1}}
\end{figure}

In classical convolutional autoencoders, both the encoder and the decoder are \textit{Convolutional Neural Networks} (\textit{CNN})~\cite{Fukushima1980}. The convolution operations allow to identify the key features in an image, thus making them well suited for classification, denoising and image compression tasks. However, since there is no any direct control on how the input space is mapped to the \textit{latent space}, it is difficult to extract specific and valuable information from the encoded \textit{semantic code vectors}.\\
One way to overcome this limitation is to replace the decoder \textit{CNN} with a given function that produces a synthetic model of the input data, as already proposed by Aragon-Calvo~\cite{Aragon-Calvo2019}. In this way, after a successful training, the \textit{latent space} is forced to coincide with the domain of the model function and each parameter of the \textit{semantic code} controls a different characteristic of the generated model, thus the name \textit{Disentangled Convolutional Autoencoder}. An interesting feature of this type of autoencoders, implicitly deriving from its construction, is that they can successfully represent only objects compatible with the model assumed. Identifying those objects means to detect artifacts, images containing wrong data, but also interesting outliers.\\
In our experiments we developed a multi-GPU \textit{DCA}, using the Python bindings of \textit{TensorFlow}~\cite{tensorflow2015-whitepaper} and its built-in Keras module~\cite{chollet2015keras}. The encoder part is made by three convolutional blocks, each one containing two convolution layers, using a ReLU activation function and followed by a $2 \times 2$ max-pooling layers (Fig. \ref{fig:enc_1}). The convolutional layers in the three blocks have respectively $32$, $64$ and $128$ kernels of size $4 \times 4$. The last max-pooling layer has $128$ channels of size $ \frac{imh}{8} \times \frac{imw}{8} $, where $imh$ and $imw$ are, respectively, height and width of the input images. This hierarchical module is then flattened and fed to a fully connected \textit{Multi-Layer Perceptron} (\textit{MLP})~\cite{VanDerMalsburg1986}, with two hidden layers of $64$ and $32$ neurons, respectively. The output layer of the MLP section has as many neurons as many parameters there are in the model used by the decoder.

The decoder is a custom TensorFlow layer object that encapsulates a given model and passes to it the encoder output $p$ along with a pair of coordinate meshes $X$ and $Y$. If needed, the model may also takes care of applying some appropriate constrains to the parameters. The coordinate meshes have the same size of the input images and contain, respectively, the $x$ and $y$ pixel coordinates.\\

\subsection{Validation with synthetic data}
\label{subsec:3_1}
In order to evaluate the autoencoder performance, we have first created two simple models of galaxy surface brightness: an exponential and a S\'{e}rsic profile model. Then we added a third more complex Bulge/Disk model, which is a linear combination of these two. For each profile we generate a set of synthetic images as decribed in Sec.~\ref{sec:2} and used half of them as train set, while the rest as blind test set.

%
%
\subsubsection{Exponential profile of galaxy surface brightness}
\label{subsubsec:3_1_1}
The exponential profile usually well describes the light distribution of the disk of a galaxy as function of the distance from its centre~\cite{binney2008}. The model we implemented has five parameters:
\begin{itemize}
\item{\textbf{$x_0$}: the $x$ coordinate of the center of the galaxy;}
\item{\textbf{$y_0$}: the $y$ coordinate of the center of the galaxy;}
\item{\textbf{$a$}: the size of the semi major axis in pixels;}
\item{\textbf{$q$}: the ratio between the minor ad major axis;}
\item{\textbf{$\theta$}: the rotation angle, defined as the angle that the major axis forms with the $x$ axis of the image.}
\end{itemize}

Using these parameters we first apply a coordinate transformation to take in account the translation and rotation of the galaxy (Eq. 1).

\begin{equation}\label{eq:2_1_coords}
\begin{aligned}
x'(x, y) &= (x - x_0) \cdot \cos \left( \theta \right) -  (y - y_0) \cdot \sin \left( \theta \right) \\
y'(x, y) &= (x - x_0) \cdot \cos \left( \theta \right) +  (y - y_0) \cdot \sin \left( \theta \right)
\end{aligned}
\end{equation}

Using the transformed coordinates we then compute the radius value for a give pixel coordinate $(x, y)$ with the eq. 2.

\begin{equation}\label{eq:2_1_radius}
r'(x, y) = \frac{1}{a} \cdot \sqrt{x'(x, y)^2 + \left( \frac{y'(x, y)}{q} \right)^2}
\end{equation}

And finally we compute the exponential brightness profile (eq. 3).

\begin{equation}\label{eq:exp_profile}
f_{exp}(x, y) = \exp(-r'(x, y))
\end{equation}

\begin{figure}[t]
\centering
\includegraphics[width=10cm]{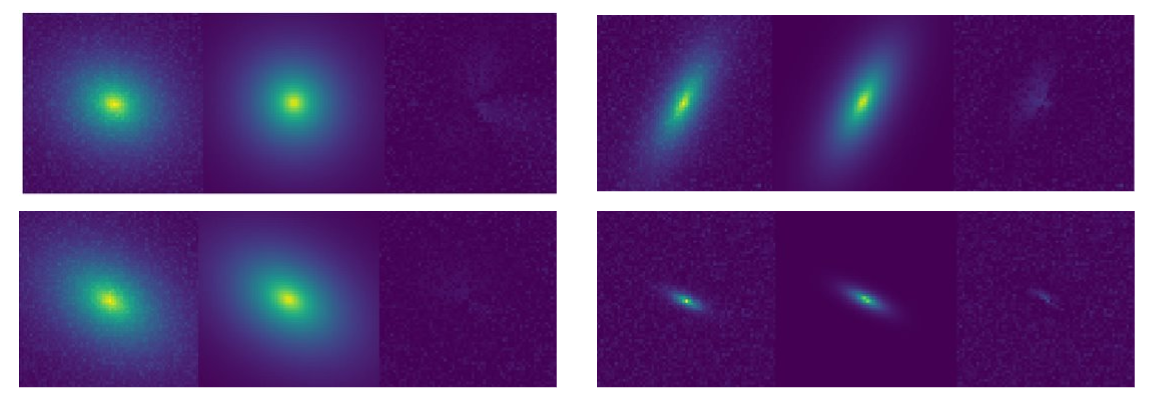}
\caption{Comparison between the autoencoder input and output images. The colour-map was chosen for a better visualization, so that the brightest pixels are yellow and the darkest ones are in dark blue. The two top images are training examples, while the two on the bottom are examples of test set. For each image, from left to right panels respectively, there are the original synthetic image with noise, the output of the autoencoder and the residual of the two images.}\label{fig:exp_example_1}
\end{figure}

This profile is normalised so that the maximum value is $1$ at $(x=x_0, y=y_0)$ and the minimum value is zero. With this profile and using random parameters we generated $20,000$ synthetic images of $64 \times 64$ pixels, according to the procedure described in Sec.~\ref{sec:2}. We then split the images into a train set and test set of $10,000$ images each. We run the autoencoder on the train set using different optimizers and loss functions. We obtained the best results using the \textit{Adam} optimizer~\cite{Kingma2015} with a learning rate of $lr=1e-4$, a batch size of $128$ images and a maximum number of $2000$ training epochs. We also used a custom loss function defined as follows (eq. 4):

\begin{equation}\label{eq:custom_loss}
\text{loss}_{mael} = \frac{1}{N \cdot \text{W} \cdot \text{H}} \sum_{j=0}^{N} \sum_{x=0,y=0}^{\text{W},\text{H}}|ln(1 + f_j(x, y)) - ln(1 + I_j(x, y))|
\end{equation}
where W and H are, respectively, the width and height of the input images $I_j$; $f_j$ is the output image generated by the autoencoder for the corresponding input image and $N$ is the total number of the images. The logarithmic transformations in eq. (\ref{eq:custom_loss}) give more weight to the fainter regions of the galaxies that are also the parts more difficult to fit. Using a higher learning rate, the training time decreases, but it also increases the chances that the algorithm will not converge to an optimal solution. Other optimizers like \textit{Adadelta}~\cite{Zeiler2012} or \textit{Stochastic gradient descent} (\textit{SGD})~\cite{Kiefer1952,Robbins1951}  very often did not converge to an optimal solution, even using different learning rates. Using these training parameters we performed $25$ executions and selected the trained model that provided the minimum \textit{mean absolute error} (\textit{MAE}) between the input and the output images and run it on the test set (Fig. \ref{fig:exp_example_1}).\\
As described in Sec.~\ref{sec:3}, the output of the autoencoder is a reconstruction of the input images, based on the parameters of the model. Therefore, to assess the goodness of the reconstructed image and in turn of the parameters we computed the \textit{MAE} and the \textit{normalised median absolute deviation} (\textit{NMAD}) of the residuals for each pair of input-output images, finding an average $\overline{MAE}=0.07 \pm 0.02$, which is compatible with the mean noise level and an average $\overline{NMAD}= 0.03 \pm 0.01$, from which we can compute the equivalent standard deviation $\overline{\sigma_{NMAD}} \approx 1.5 \cdot \overline{NMAD} = 0.05 \pm 0.02$, which is compatible with the standard deviation of the noise.
In Table \ref{tbl:exp_stasts} the normalised \textit{MAE} and \textit{NMAD} for each parameter of the model are also reported, computed using the true parameter values and the ones predicted by the trained encoder. The small values of these statistical indicators show that the autoencoder was able to successfully train the model.\\

\begin{table}
\centering
\begin{tabular}{ c | c c c c c}
parameter & $x_0$ & $y_0$ & $a$ & $q$ & $\theta$ \\
\hline\hline
NMAE & 0.16 & 0.16 & 0.32 & 0.02 & 0.01\\
NMAD & 0.02 & 0.02 & 0.03 & 0.01 & 0.01\\
\end{tabular}
\caption{Statistical estimators for the true vs. predicted values for each parameter of the exponential galaxy profile model. Note that, although the uncertainty on the size of the galaxy is relatively larger than other parameters, the uncertainty on the axis ratio is small.\label{tbl:exp_stasts}}
\end{table}

%
%
\subsubsection{S\'{e}rsic profile of galaxy surface brightness}
\label{subsubsec:3_1_2}
The S\'{e}rsic profile usually describes well the light distribution of the bulge of a galaxy and elliptical galaxies in general~\cite{Graham2005,Roy2018}. This profile is a generalization of the exponential profile, obtained by introducing a parameter $n$, called \textit{S\'{e}rsic index}, that controls how the light is distributed across the galaxy. The inverse of this parameter $\beta=1/n$ is used as exponent of the radius in the surface brightness profile equation. For $n=1$ the S\'{e}rsic profile coincides with the exponential one. The model we implemented has six parameters:
\begin{itemize}
\item{\textbf{$x_0$}: the $x$ coordinate of the center of the galaxy;}
\item{\textbf{$y_0$}: the $y$ coordinate of the center of the galaxy;}
\item{\textbf{$a$}: the size of the semi major axis in pixels;}
\item{\textbf{$c$}: the ratio between the minor ad major axis;}
\item{\textbf{$\theta$}: the rotation angle defined as the angle that the major axis forms with the $x$ axis of the image;}
\item{\textbf{$\beta$}: the inverse of the S\'{e}rsic index.}
\end{itemize}

The only difference with the exponential model is in the profile function of eq. (\ref{eq:exp_profile}) that becomes as specified in eq. (5).
\begin{equation}\label{eq:sersic_profile}
f_{sersic}(x, y) = \exp(-{r'(x, y)}^{\beta})
\end{equation}

As in the previous profile case, by using random parameters we generate $20,000$ synthetic images, divided into a training and a test set, each one composed by $10,000$ images.
Also in this case we obtained the best training results using the \textit{Adam} optimizer with the custom loss function (\ref{eq:custom_loss}), a learning rate of $lr=1e-4$, a batch size of $128$ images, and a maximum number of training epochs of $2000$. We computed the \textit{MAE} and the \textit{NMAD} of the residuals for each pair of input-output images, finding an average $\overline{MAE}=0.06 \pm 0.03$ that is compatible with the mean noise level and an average $\overline{NMAD}= 0.03 \pm 0.01$. In Table \ref{tbl:sersic_stasts} the normalised \textit{MAE} and \textit{NMAD} for each parameter of the model are reported, computed using the true parameter values and the ones predicted by the trained encoder. The small values of these statistical estimators show that the autoencoder was able to successfully train the model.\\

\begin{table}
\centering
\begin{tabular}{ c | c c c c c c}
parameter & $x_0$ & $y_0$ & $a$ & $q$ & $\theta$ & $\beta$\\
\hline\hline
NMAE & 0.17 & 0.17 & 0.35 & 0.02 & 0.02 & 0.33\\
NMAD & 0.02 & 0.02 & 0.04 & 0.01 & 0.01 & 0.08\\
\end{tabular}
\caption{Statistical estimators for the true vs. predicted values, reported for each parameter of the S\'{e}rsic galaxy profile model.\label{tbl:sersic_stasts}}
\end{table}

%
%
\subsubsection{Bulge/Disk profile of galaxy surface brightness}
\label{subsubsec:3_1_3}
This model is a linear combination of the Exponential and S\'{e}rsic profiles, used to mimic a combination of bulge and disk components as well as a uniform background. We introduced also a constant background level to take into account the sky background present in almost all real images. It has eleven parameters:
\begin{itemize}
\item{\textbf{$x_0$}: the $x$ coordinate of the center of the galaxy;}
\item{\textbf{$y_0$}: the $y$ coordinate of the center of the galaxy;}
\item{\textbf{$a_{disk}$}: the size of the semi major axis of the disk component in pixels;}
\item{\textbf{$c_{disk}$}: the ratio between the minor ad major axis of the disk component;}
\item{\textbf{$\theta_{disk}$}: the rotation angle of the disk component, defined as the angle that the major axis forms with the $x$ axis of the image;}
\item{\textbf{$\alpha$}: the fractional ratio between the central brightness of the bulge and the central brightness of the disk;}
\item{\textbf{$a_{bulge}$}: the size of the semi major axis of the bulge component in pixels;}
\item{\textbf{$c_{bulge}$}: the ratio between the minor ad major axis of the bulge component;}
\item{\textbf{$\theta_{bulge}$}: the rotation angle of the bulge component, defined as the angle that the major axis forms with the $x$ axis of the image;}
\item{\textbf{$\beta$}: the inverse of the bulge S\'{e}rsic index;}
\item{\textbf{$k$}: the background level expressed as fractional ratio between the brightness of the background and the maximum brightness of bulge+disk.}
\end{itemize}

The profile function of this model is shown in the eq. (6).
\begin{equation}\label{eq:bd_profile}
f_{bd}(x, y) = (1 - k) \cdot \left( \alpha \cdot f_{sersic}(x, y) + (1 - \alpha) \cdot f_{exp}(x, y) \right) + k
\end{equation}

As done in the previous tests, by using random parameters, we generate $20,000$ synthetic images, divided into a training and a test set, each one containing $10,000$ images. Also in this case we obtained the best training results using the Adam optimizer with the custom loss function (\ref{eq:custom_loss}), a learning rate of $lr=1e-4$, a batch size of $128$ images, and a maximum number of $2000$ training epochs. The results of the test and the training were similar to those found in the previous test, finding an average $\overline{MAE}=0.07 \pm 0.04$ that is compatible with the mean noise level and an average $\overline{NMAD}= 0.03 \pm 0.02$.

\subsection{Application to KiDS data}
\label{subsec:3_2}
After having validated the autoencoder model on synthetic data, we tried to apply the \textit{Bulge/Disk} profile model on real data. As already introduced in Sec. \ref{sec:2}, the images used in this experiment are cutouts taken from the $r$ band tiles of the KIDS DR4. We divided them into a training set of $30,000$ images and a set of $370,000$ images used to detect potentially interesting outliers.\\
We trained the autoencoder by using the \textit{Bulge/Disk} profile model, the optimizer and training parameters validated with synthetic data. $25$ training runs were performed, selecting the trained model with the lowest \textit{MAE}. Finally we run the best trained model on the image test set.

\subsubsection{Anomaly detection with DCA}
\label{subsubsection:2_2_1}
As we said above, if the autoencoder is correctly trained and the chosen model is a valid representation of the input objects, then the residual images - obtained by subtracting the output of the decoder from the corresponding input - should contain only residual noise. Therefore, it is clear that the statistical estimators computed on the residual images have a key role in detecting anomalies that the model is not able to describe. We used the following statistical estimators:

\begin{itemize}
    \item{\textbf{MAD}: since it is not very influenced by extreme values, the median of the pixel values in the residual image corresponds approximately to the mean background value. Thus, the \textit{Median Absolute Deviation} is a valid measure of how broadly the residuals are distributed around the background. A high value could indicate the presence of substructures or artifacts.}
    \item{\textbf{Skewness}: unusually high or low values of this statistical moment could indicate that there is something odd in the image.}
    \item{\textbf{Maximum}: hot pixels, artifacts but also other objects in the whole image produce very bright pixels in the residual images.}
\end{itemize}

\begin{figure}
    \centering
    \begin{subfigure}[b]{.9\textwidth}
        \centering
        \includegraphics[width=\linewidth]{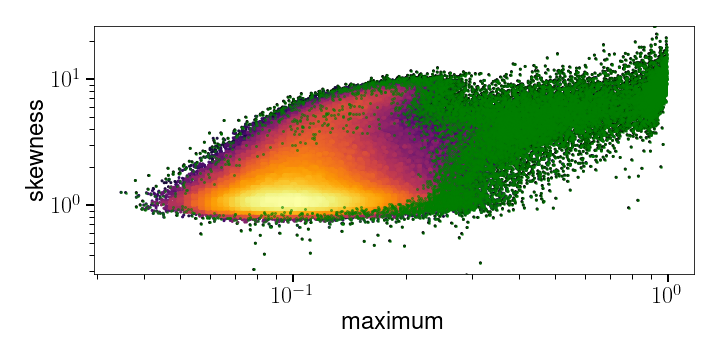}
    \end{subfigure}
    \quad
    \begin{subfigure}[b]{.9\textwidth}
        \centering
        \includegraphics[width=\linewidth]{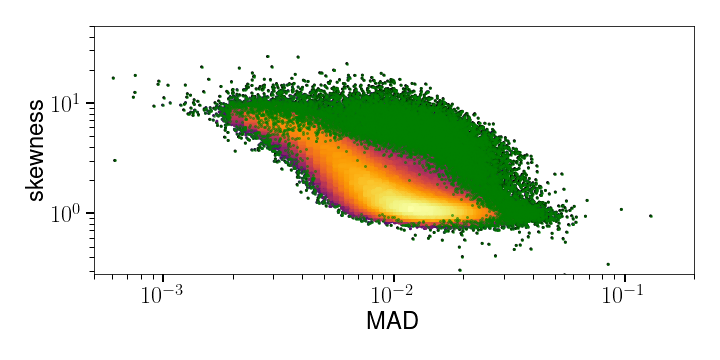}
    \end{subfigure}
    \quad
    \begin{subfigure}[b]{.9\textwidth}
        \centering
        \includegraphics[width=\linewidth]{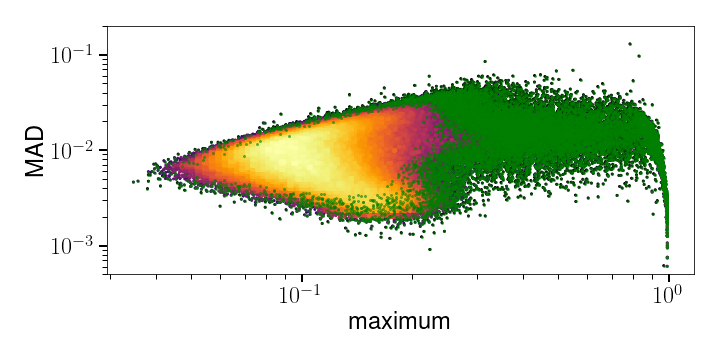}
    \end{subfigure}
    \caption{From top to bottom: scatter plots of the skewness vs. maximum, skewness vs. MAD and MAD vs maximum. Axes are in logarithmic scale and the colour indicates the logarithm of the local number density of the points, where a lighter colour means a denser region. Objects identified as outliers are highlighted in green.\label{fig:stats_dca}}
\end{figure}

\begin{figure}
    \centering
    \includegraphics[width=\linewidth]{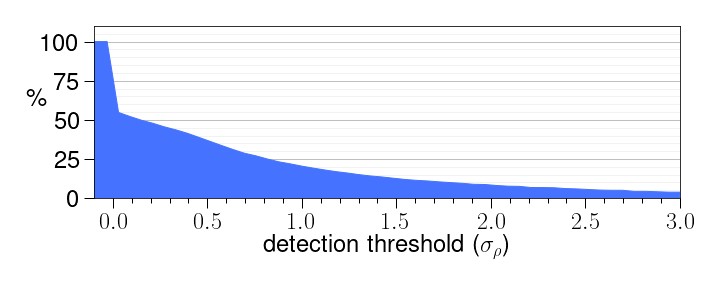}
    \caption{Percentages of objects classified as outliers by the DCA, as a function of the detection threshold expressed in units of $\sigma_{\rho}$.}
    \label{fig:dca_cumulative_sigma}
\end{figure}

The outliers were selected using the following automated procedure: as first step, the average maximum value of the residuals $\overline{max}$ was computed along with the respective standard deviation $\sigma_{max}$ and all objects for which $max > \overline{max} + 3\sigma_{max}$ were marked has outliers. Then for each unique pair of statistical estimators, the average number density of the objects, $\overline{\rho_n}$, and the corresponding standard deviation $\sigma_{n}$ were computed; then this two-dimensional space was divided into $400$ tiles of equal size. The local number density $\rho_n$ was computed in each tile and the resulting density map was smoothed with a gaussian kernel. Finally, each object falling in a sub-region with a density $\rho < \overline{\rho_n} - 2\sigma_{\rho}$ was marked as outlier, as it can be seen in Fig. \ref{fig:stats_dca}. In Fig.~\ref{fig:dca_cumulative_sigma} the percentage of objects detected as outliers is reported as a function of the detection threshold previously defined. We note the robustness of the detection that remains approximately constant above the value of~$2\sigma_{\rho}$. About the $93\%$ of the objects are concentrated in a quite continuous region with an average of~$\overline{MAD} = 0.011 \pm 0.005$,~$\overline{skewness} =  2.3 \pm 1.7$~and~$\overline{max} = 0.12 \pm 0.03$. These are objects that the autoencoder was able to fit with the model. The low values of the $\overline{MAD}$ and $\overline{max}$ indicate that the reconstructions of the autoencoder describe very well the original images and that the residuals contain basically only background noise. This is confirmed also by the value of $\overline{skewness}$ greater than zero, which is typical of Poissonian distributions, characterized by a low value of the mean, as in the case of the shot noise that affects digital images.
We identified few objects having a very low skewness, which usually indicates a Gaussian-like distribution of the residuals and thus the presence of something else beyond the pure poissonian background noise. 

\begin{figure}
\centering
\includegraphics[width=\linewidth]{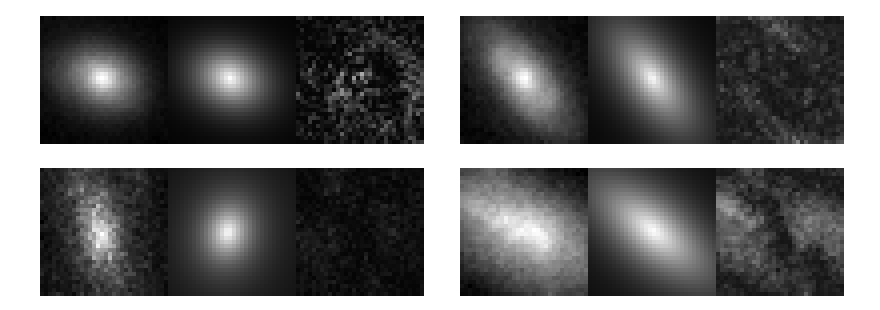}
\caption{Some of these objects show the presence of substructures that were hidden by the light of the galaxy (upper left and upper right), while in other cases the autoencoder failed to fit the surface brightness profile, because the objects were bigger than the size of the cutout. For each image, from left to right panels, there is the original image, the images produced by the autoencoder and the residual image, re-scaled to highlight the presence of substructures.\label{fig:out_skw}}
\end{figure}

\begin{figure}
\centering
\includegraphics[width=\linewidth]{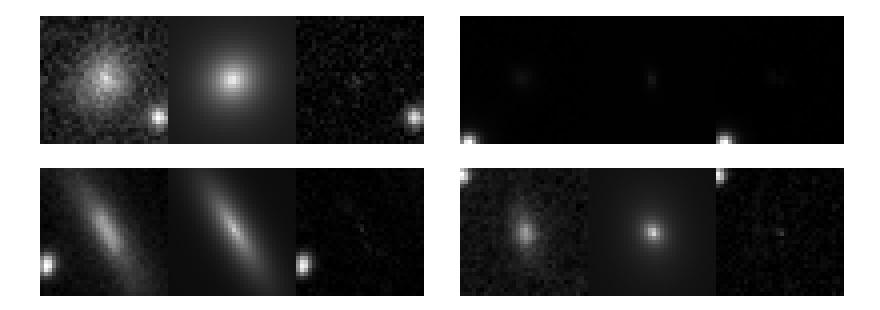}
\caption{Examples of KiDS galaxy that are very faint or have a very bright close companion or present artifacts like hot-pixels. For each image, from left to right panels, there is the original image, the images produced by the autoencoder and the residual image.\label{fig:out_max}}
\end{figure}

Some of these objects were bigger than the cutout area (Fig.~\ref{fig:out_skw}), which the autoencoder was less able to fit properly, while others showed traces of substructures in the residual image, which were hidden by the galaxy light. There was also a small clump of objects, less than $2\%$ of the total amount, having a very high maximum value: these were very faint sources or objects with a very bright companion (Fig.~\ref{fig:out_max}). 

\begin{figure}
\centering
\includegraphics[width=\linewidth]{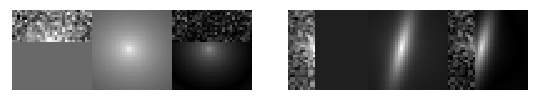}
\caption{Examples of objects that are just on the border of the tile from where the cutouts have been extracted. For each image, from left to right panels, there is the original image, the images produced by the autoencoder and the residual image.\label{fig:out_max_mad}}
\end{figure}

\begin{figure}
\centering
\includegraphics[width=\linewidth]{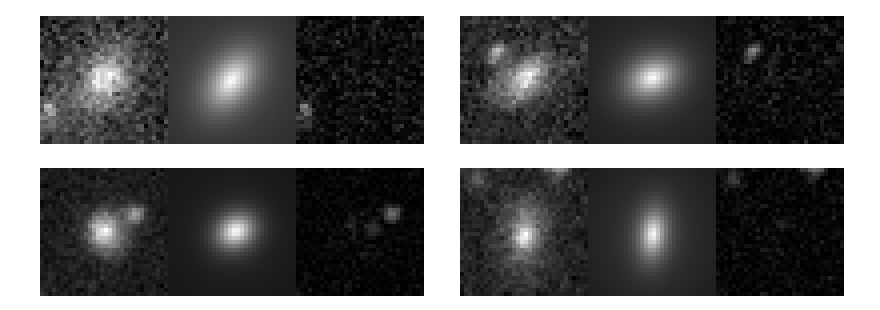}
\caption{Examples of objects showing the presence of a faint quite close companion. For each image, from left to right panels, there is the original image, the images produced by the autoencoder and the residual image.\label{fig:out_brg}}
\end{figure}

A small set of objects have also a very low \textit{MAD}. Although a low value of this statistical estimator could imply a low dispersion of the residuals, an unusual low value means that most of the pixels in the residual image have the same value, which in turn could indicate some sort of corruption. In fact, most of these objects were located on the edge of the tile, thus resulting in a partially corrupted cutouts (Fig.~\ref{fig:out_max_mad}). Finally, there was a subset of objects, approximately the $5\%$ of the total amount, having a fairly average of \textit{MAD} and \textit{skewness} values, but with a quite large maximum value ranging from $0.3$ to $0.8$. Almost all of these objects have one or more than one faint companion, as shown in the examples of Fig.~\ref{fig:out_brg}. 

\section{Unsupervised Random Forests}
\label{sec:4}

Random Forests are an ensemble of several independently grown decision tree classifiers, where each tree is a non-parametric model organized in a top-bottom tree-like structure and is grown using a random subset of the features of the training dataset~\cite{Breiman2001}. They are usually used to classify objects for which a training set of labeled objects exists, so that each tree in the forest learns to map the input features to the corresponding correct label. When a object identified by a set of feature is passed to the forest, each tree votes for its belonging to one of the given classes, identified by the labels, and the resulting class is usually determined by majority voting.\\
For the problem of outliers detection, where obviously a labelled training set it not available, random forests can also be used in an unsupervised configuration.\\ 
A simple but efficient way to use Random Forest as an unsupervised method, is to generate a synthetic dataset from the original one, with same size and same marginal distribution in all its features, but without the covariance among objects. Then the Random Forest is trained on both datasets to learn to recognize their similarity, thus isolating the outliers. By defining a \textit{similarity index} $S_{i,j}$ between any two objects as the number of common "real" leaves of the trees, divided by the total number of trees in the forest, a \textit{weirdness score} can be introduced, which describes how distant is, on average, from all the others. This score can assume any value between 0 and 1, but the distribution of its values mostly depends on the specific dataset involved. Therefore, a reasonable way to use it is to impose a certain threshold, based on the distribution of its values for all the objects in the dataset and then to consider as outliers all objects with a weirdness value greater of such threshold. \textit{Baron \& Poznanski}~\cite{Baron2017} proposed this method that was able to find some galaxies with peculiar spectra in the 12th data release of the Sloan Digital Sky Survey~\cite{Alam2015}.

\subsection{Anomaly detection in KiDS data based on the URF}
\label{subsec:4_2}

\begin{figure}
\centering
\includegraphics[width=\linewidth]{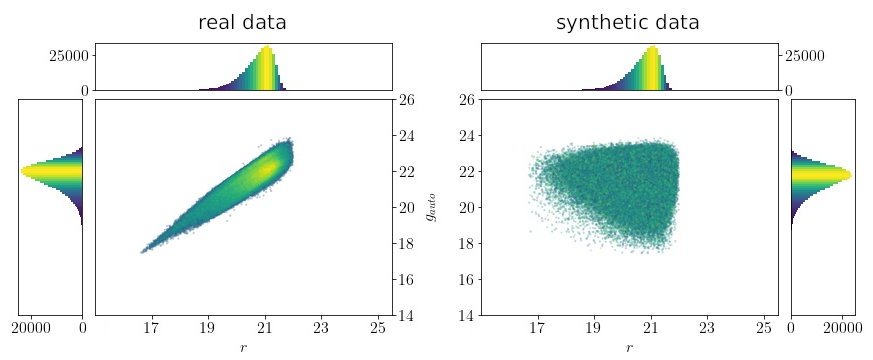}
\caption{Projection on two features (magnitudes r and g) of the density distributions of the KiDS sample (left panel) and synthetic data (right panel). The latter was generated from the same marginal distributions of the real one, by removing the covariance among original data.\label{fig:marg_dist_example}}
\end{figure}

For this experiment we used the photometric catalogue containing the counterparts of the image cutouts, organized as described in Sec.~\ref{sec:2}. According to what described in Sec.~\ref{sec:4}, we then created a synthetic dataset of the same size of the real one and with objects drawn randomly from the same marginal distribution of each feature (Fig. \ref{fig:marg_dist_example}).

\begin{figure}
\centering
\includegraphics[width=\linewidth]{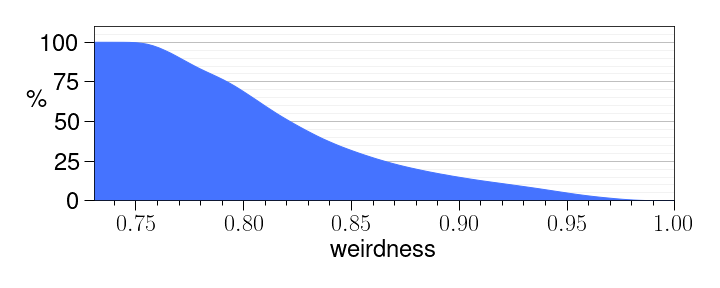}
\caption{Percentages of objects classified as outliers by the \textit{URF}, as a function of the weirdness.}\label{fig:urf_cumulative_w}
\end{figure}

We then merged the two datasets into a single one, labelling the objects depending on whether they were real or synthetic, and used for training and testing a \emph{Random Forest Classifier} containing $800$ trees, built using the Python package \textit{scikit-learn}~\cite{scikit-learn}. We remark that in the case of the \textit{URF} model, both training and testing sets coincide, since same data are used to perform the anomaly detection experiment along the construction of the random forest trees process.
We then divided the original dataset in batches of $6000$ objects and computed the \textit{weirdness} index for each batches. The size was limited by the amount of memory necessary to compute the \textit{weirdness}. The whole process has been repeated $4$ times and the \textit{weirdness} values have been averaged for each object. The objects show a distribution centered on an average value of \textit{weirdness} of $\overline{W}=0.83$ with a standard deviation of $\sigma_{w} = 0.06$, while the number of objects decreases as the \textit{weirdness} value increases (Fig. \ref{fig:urf_cumulative_w}). In analogy to what done in the case of the \textit{DCA} model (Sec. \ref{subsubsection:2_2_1}), and to perform a direct comparison between the two models, we imposed a detection threshold of $2\sigma$ and considered as outliers all objects for which $w > \overline{w} + 2\sigma_{w} = 0.95$.

\section{Discussion}
\label{sec:5}

\begin{figure}
    \centering
    \begin{subfigure}[b]{0.7\textwidth}
        \centering
        \includegraphics[width=\linewidth]{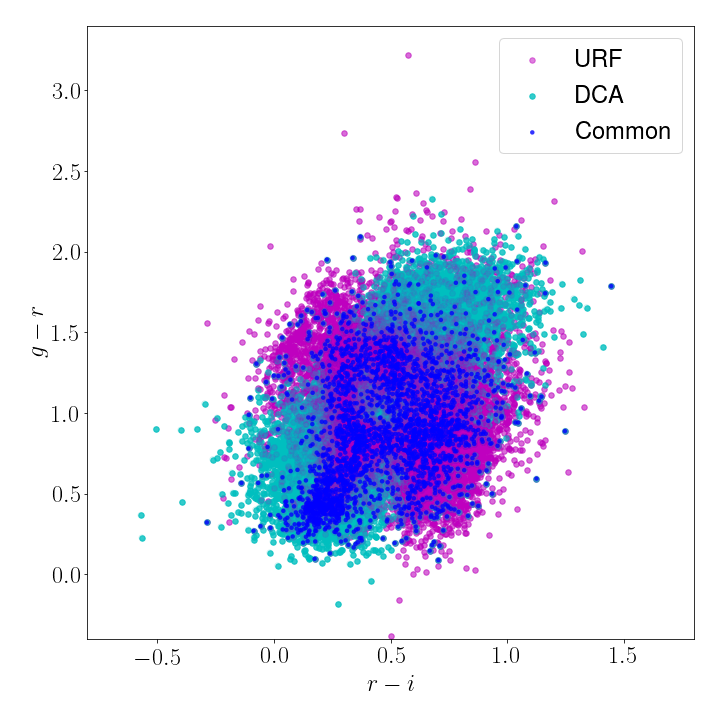}
        \caption{\label{fig:outl_gi_ri_com}}    
    \end{subfigure}
    \quad
    \begin{subfigure}[b]{0.7\textwidth}
        \centering
        \includegraphics[width=\linewidth]{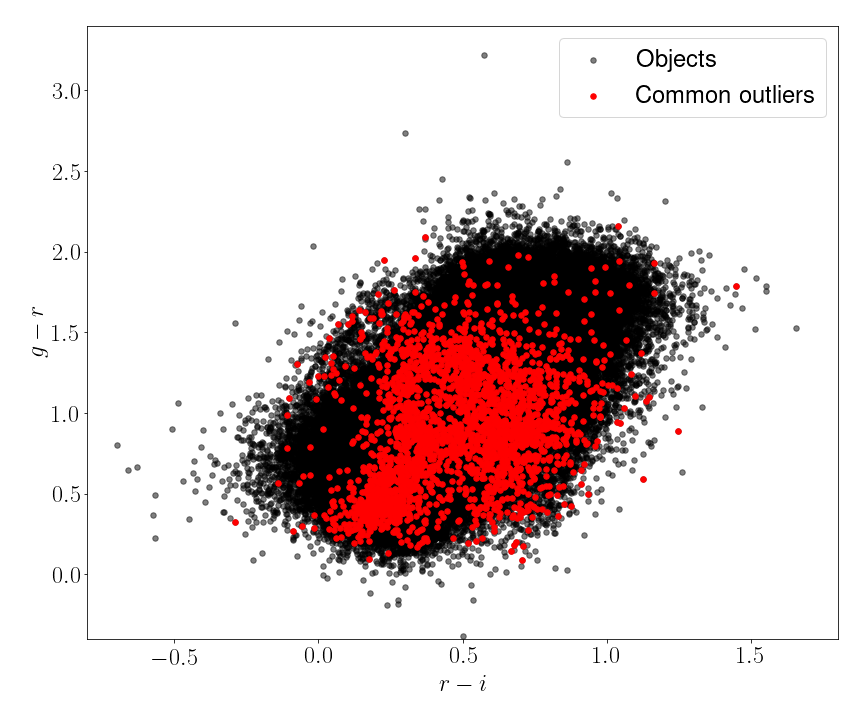}
        \caption{\label{fig:out_tot_com_u}}
    \end{subfigure}
    \caption{Upper panel: colour-colour diagram of all the candidate outliers detected by \textit{URF} (in magenta) and \textit{DCA} (in cyan). Common outliers found by both methods are coloured in blue. Lower panel: colour-colour diagram of the outliers detected by both \textit{DCA} and \textit{URF} (in red), plotted against all the objects in the dataset (in black).
    \label{fig:outl_com}}
\end{figure}

\begin{figure}
    \centering
    \begin{subfigure}[b]{0.24\textwidth}
        \includegraphics[width=\linewidth]{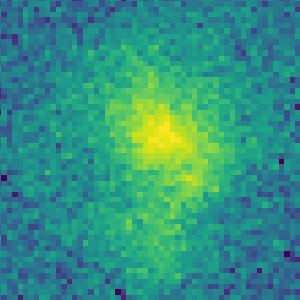}
        \caption{\label{fig:outl_irr_1}}
    \end{subfigure}
    \begin{subfigure}[b]{0.24\textwidth}
        \includegraphics[width=\linewidth]{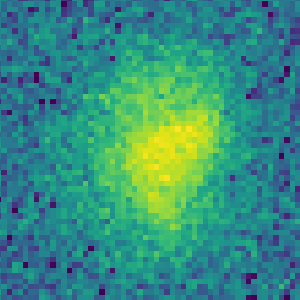}
        \caption{\label{fig:outl_irr_2}}
    \end{subfigure}
    \begin{subfigure}[b]{0.24\textwidth}
        \includegraphics[width=\linewidth]{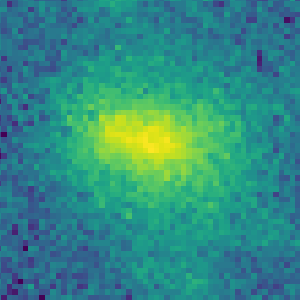}
        \caption{\label{fig:outl_irr_3}}
    \end{subfigure}
    \begin{subfigure}[b]{0.24\textwidth}
        \includegraphics[width=\linewidth]{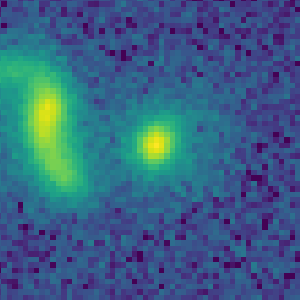}
        \caption{\label{fig:outl_merg_1}}
    \end{subfigure}
    \quad
    \caption{Examples of cotouts including irregular galaxies (a, b and c) and interacting galaxies (d), detected as anomalies by the two methods.}
    \label{fig:outl_exmpl_irr}
\end{figure}

\begin{figure}
    \centering
    \begin{subfigure}[b]{0.24\textwidth}
        \includegraphics[width=\linewidth]{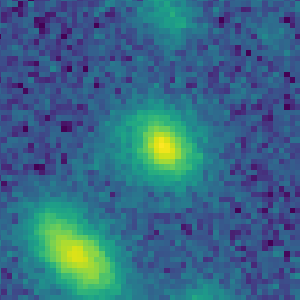}
        \caption{\label{fig:outl_cf_1}}
    \end{subfigure}
    \begin{subfigure}[b]{0.24\textwidth}
        \includegraphics[width=\linewidth]{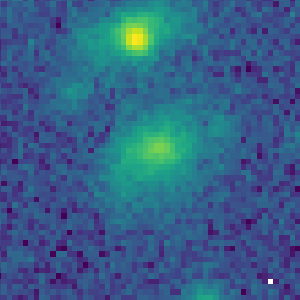}
        \caption{\label{fig:outl_cf_2}}
    \end{subfigure}
    \begin{subfigure}[b]{0.24\textwidth}
        \includegraphics[width=\linewidth]{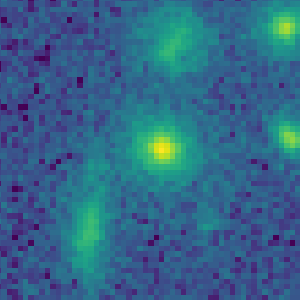}
        \caption{\label{fig:outl_cf_3}}
    \end{subfigure}
    \begin{subfigure}[b]{0.24\textwidth}
        \includegraphics[width=\linewidth]{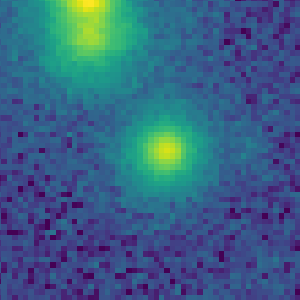}
        \caption{\label{fig:outl_cf_4}}
    \end{subfigure}
    \quad
    \begin{subfigure}[b]{0.24\textwidth}
        \includegraphics[width=\linewidth]{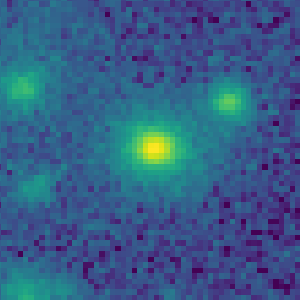}
        \caption{\label{fig:outl_cf_5}}
    \end{subfigure}
    \begin{subfigure}[b]{0.24\textwidth}
        \includegraphics[width=\linewidth]{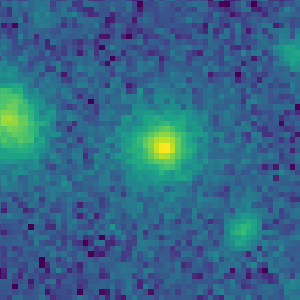}
        \caption{\label{fig:outl_cf_6}}
    \end{subfigure}
    \begin{subfigure}[b]{0.24\textwidth}
        \includegraphics[width=\linewidth]{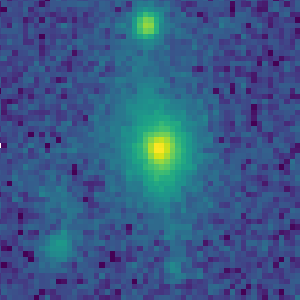}
        \caption{\label{fig:outl_cf_7}}
    \end{subfigure}
    \begin{subfigure}[b]{0.24\textwidth}
        \includegraphics[width=\linewidth]{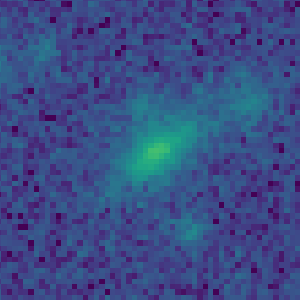}
        \caption{\label{fig:outl_cf_8}}
    \end{subfigure}
    \caption{Examples of cutouts for sources within crowded fields, detected as anomalies by the two methods.}
    \label{fig:outl_exmpl_cfs}
\end{figure}

Both chosen algorithms, \emph{DCA} and \emph{URF}, were tested on a subset of $400,000$ objects extracted from the KiDS survey Data Release 4. The \emph{DCA} was used directly on image cutouts extracted from the \textit{r} band coadds tiles, while the \textit{URF} was used on the catalogue of counterparts, made by magnitudes in the bands \textit{ugri}, their derived colours and ratios. The model \textit{DCA} required only a minimum of human supervision during the training, just to check the convergence of the algorithm to an optimal solution. It performed very well in both terms of memory requirements and computing time and was able to pinpoint some peculiar sources, about the $5$\% of the sample, showing substructures that were hidden by the close galaxy light, as well as objects with very small and/or faint close companions.\\
Since the \textit{URF} is based on the computation of a \emph{similarity matrix}, whose size increases as the square of the number of the objects, the dataset has to be analysed in batches and a supplementary amount of human intervention was required in order to determine the optimal batch size.\\
To perform a comparative analysis of the results obtained by the two methods, we imposed a similar criterion to extract candidate outliers, for instance, a common value of $2\sigma$ with $w > \overline{w} + 2\sigma_{w} = 0.95$ in terms of weirdness $w$ for the URF and object density $\rho < \overline{\rho_n} - 2\sigma_{\rho}$ in the case of DCA.
With such outlier detection thresholds, both methods found a comparable amount of peculiar objects, $\sim 7\%$ of the test set for \textit{DCA} and $\sim 5$ for URF. Among the objects considered as an anomaly by at least one of the two methods, the $\sim 7\%$ were detected as peculiar objects by both of them. The distributions of the outliers (Fig.~\ref{fig:outl_com}) shows that most of the peculiar objects found by the two models cover a wide and uncorrelated area of the parameter space, with a limited overlapping region in which most of the common outliers lay. This seems to suggest a certain amount of complementarity of the two methods in detecting peculiarities, according to a similar behaviour found in~\cite{brescia2019}, concerning the analysis of outliers identified from a distribution of photometric redshifts, estimated by different methods, however no any particular evidence of interesting peculiarity seems to emerge.\\
By analyzing the detected peculiar objects having a class label provided in \cite{Nakoneczny2019}, only about the $\sim27\%$ of stars and QSOs were detected as anomalies. 
Most of these objects, in fact, were not confirmed as peculiar by \textit{DCA} and appear uniformly distributed with respect to the different thresholds of weirdness calculated by the \textit{URF}. This behaviour was expected for \textit{DCA} because no any limitation was imposed on the value of the S\'{e}rsic index nor on the galaxy size, thus the model should be able to fit also star-like objects. Through a visual inspection of the cutouts for the peculiar objects detected, we observe that both methods tend to assign as peculiar the irregular and interacting galaxies (see examples in Fig. \ref{fig:outl_exmpl_irr}), as well as objects that are in more crowded fields, like the ones showed in Fig. \ref{fig:outl_exmpl_cfs}.\\ 

\section{Conclusions}
\label{sec:6}
The identification of anomalies in Astronomy has always played a major role in making new scientific discoveries. Nowadays, the shift to more large and complex surveys makes essential the use of robust and efficient automated algorithms to identify peculiar patterns. In this context we performed a preliminary set of anomaly detection experiments, by testing two different unsupervised machine learning algorithms, a \emph{Disentangled Convolutional Autoencoder} and an \emph{Unsupervised Random Forest}, using the former on real image cutouts and the latter on the catalogue of their counterparts, which includes measured magnitudes, derived colours and magnitude ratios, both extracted from the 4th KiDS Data Release.\\
We performed a comparative analysis of the peculiar objects detected by both methods, by analyzing their colour distribution in the parameter space and their capability to disentangle the presence of QSOs and stars from galaxies within a mixed datasets. The results of this preliminary experiment revealed that most of the anomalies detected by both methods involve irregular and interacting galaxies and sources located in more crowded fields. Further experiments are then required on these models, especially in terms of their setup and configuration, to investigate their real capability to isolate peculiar types of sources. In particular, since the \textit{DCA} is mainly a method to estimate the goodness of a fit to the data, it may result affected by the presence of nearby objects, not taken into account by the model. Improving the detection criteria for \textit{DCA} is thus one of the future enhancements of this method, as well as to take into account the PSF and seeing in the Bulge/Disk model, which should achieve a more accurate estimation of the structural parameters. Regarding \textit{URF}, on the other hand, a further step is the introduction of the infrared bands in the photometric dataset, as well as the search for spectroscopic counterparts, which can improve the classification accuracy and the validation of the method.

\begin{acknowledgement}
Based on observations made with ESO Telescopes at the La Silla Paranal Observatory under programme IDs 177.A-3016, 177.A-3017, 177.A-3018 and 179.A-2004, and on data products produced by the KiDS consortium. The KiDS production team acknowledges support from: Deutsche Forschungsgemeinschaft, ERC, NOVA and NWO-M grants; Target; the University of Padova, and the University Federico II (Naples).
MB acknowledges financial contributions from the agreement \textit{ASI/INAF 2018-23-HH.0, Euclid ESA mission - Phase D}. MB and CT acknowledge the \textit{INAF PRIN-SKA 2017 program 1.05.01.88.04}. SC acknowledges the financial contribution from FFABR 2017.
\end{acknowledgement}

\bibliographystyle{spphys}
\bibliography{main}{}

\begin{thebibliography}{10}
\providecommand{\url}[1]{{#1}}
\providecommand{\urlprefix}{URL }
\expandafter\ifx\csname urlstyle\endcsname\relax
  \providecommand{\doi}[1]{DOI \discretionary{}{}{}#1}\else
  \providecommand{\doi}{DOI \discretionary{}{}{}\begingroup
  \urlstyle{rm}\Url}\fi

\bibitem{Baron2017}
D.~Baron, D.~Poznanski, Monthly Notices of the Royal Astronomical Society
  \textbf{465}(4), 4530 (2017).
\newblock \doi{10.1093/mnras/stw3021}.
\newblock
  \urlprefix\url{https://academic.oup.com/mnras/article-lookup/doi/10.1093/mnras/stw3021}

\bibitem{brescia2018}
M.~{Brescia}, S.~{Cavuoti}, V.~{Amaro}, G.~{Riccio}, G.~{Angora},
  C.~{Vellucci}, G.~{Longo}, in \emph{Data Analytics and Management in Data
  Intensive Domains}, \emph{Communications in Computer and Information
  Science}, vol. 822, ed. by L.~{Kalinichenko}, Y.~{Manolopoulos}, O.~{Malkov},
  N.~{Skvortsov}, S.~{Stupnikov}, V.~{Sukhomlin} (Springer International
  Publishing, 2018), \emph{Communications in Computer and Information Science},
  vol. 822, pp. 61--72

\bibitem{fluke2019}
C.J. {Fluke}, C.~{Jacobs}, arXiv e-prints arXiv:1912.02934 (2019)

\bibitem{Shi2006}
T.~Shi, S.~Horvath, Journal of Computational and Graphical Statistics  (2006).
\newblock \doi{10.1198/106186006X94072}

\bibitem{chen2018}
R.T.Q. Chen, X.~Li, R.B. Grosse, D.K. Duvenaud, in \emph{Advances in Neural
  Information Processing Systems 31}, ed. by S.~Bengio, H.~Wallach,
  H.~Larochelle, K.~Grauman, N.~Cesa-Bianchi, R.~Garnett (Curran Associates,
  Inc., 2018), pp. 2610--2620.
\newblock
  \urlprefix\url{http://papers.nips.cc/paper/7527-isolating-sources-of-disentanglement-in-variational-autoencoders.pdf}

\bibitem{guo2017}
X.~Guo, X.~Liu, E.~Zhu, J.~Yin, in \emph{Neural Information Processing}, ed. by
  D.~Liu, S.~Xie, Y.~Li, D.~Zhao, E.S.M. El-Alfy (Springer International
  Publishing, Cham, 2017), pp. 373--382

\bibitem{Tuccillo2018}
D.~Tuccillo, M.~Huertas-Company, E.~Decenci{\`{e}}re, S.~Velasco-Forero,
  H.~{Dom{\'{i}}nguez S{\'{a}}nchez}, P.~Dimauro, Monthly Notices of the Royal
  Astronomical Society \textbf{475}(1), 894 (2018).
\newblock \doi{10.1093/mnras/stx3186}

\bibitem{Reis2018a}
I.~Reis, D.~Baron, S.~Shahaf, The Astronomical Journal  (2018).
\newblock \doi{10.3847/1538-3881/aaf101}

\bibitem{Reis2018}
I.~{Reis}, D.~{Poznanski}, P.B. {Hall}, MNRAS \textbf{480}(3), 3889 (2018).
\newblock \doi{10.1093/mnras/sty2127}

\bibitem{erdmann2019}
M.~{Erdmann}, F.~{Schl{\"u}ter}, R.~{{\v{S}}m{\'\i}da}, Journal of
  Instrumentation \textbf{14}(4), P04005 (2019).
\newblock \doi{10.1088/1748-0221/14/04/P04005}

\bibitem{Kuijken2019}
K.~Kuijken, C.~Heymans, A.~Dvornik, H.~Hildebrandt, J.T. {De Jong}, et~al.,
  Astronomy and Astrophysics  (2019).
\newblock \doi{10.1051/0004-6361/201834918}

\bibitem{dejong2017}
J.T.A. {de Jong}, G.A. {Verdoes Kleijn}, T.~{Erben}, H.~{Hildebrandt},
  K.~{Kuijken}, et~al., \aap \textbf{604}, A134 (2017).
\newblock \doi{10.1051/0004-6361/201730747}

\bibitem{disanto2018}
{D\'{}Isanto, A.}, {Cavuoti, S.}, {Gieseke, F.}, {Polsterer, K. L.}, A\&A
  \textbf{616}, arXiv:1904.07248 (2018).
\newblock \doi{10.1051/0004-6361/201833103}.
\newblock \urlprefix\url{https://doi.org/10.1051/0004-6361/201833103}

\bibitem{Nakoneczny2019}
S.~Nakoneczny, M.~Bilicki, A.~Solarz, A.~Pollo, N.~Maddox, C.~Spiniello,
  M.~Brescia, N.R. Napolitano, Astronomy and Astrophysics  (2019).
\newblock \doi{10.1051/0004-6361/201834794}

\bibitem{Goodfellow-et-al-2016}
I.~Goodfellow, Y.~Bengio, A.~Courville, \emph{Deep Learning} (MIT Press, 2016).
\newblock \url{http://www.deeplearningbook.org}

\bibitem{Fukushima1980}
K.~Fukushima, Biological Cybernetics  (1980).
\newblock \doi{10.1007/BF00344251}

\bibitem{Aragon-Calvo2019}
M.A. {Aragon-Calvo}, arXiv e-prints arXiv:1907.03957 (2019)

\bibitem{tensorflow2015-whitepaper}
M.~{Abadi}, A.~{Agarwal}, P.~{Barham}, E.~{Brevdo}, Z.~{Chen}, et~al.
\newblock {TensorFlow}: Large-scale machine learning on heterogeneous systems
  (2015).
\newblock \urlprefix\url{https://www.tensorflow.org/}.
\newblock Software available from tensorflow.org

\bibitem{chollet2015keras}
F.~Chollet, et~al.
\newblock Keras.
\newblock \url{https://keras.io} (2015)

\bibitem{VanDerMalsburg1986}
C.~{Van Der Malsburg}, in \emph{Brain Theory} (Springer Berlin Heidelberg,
  Berlin, Heidelberg, 1986), pp. 245--248.
\newblock \doi{10.1007/978-3-642-70911-1\_20}.
\newblock
  \urlprefix\url{http://link.springer.com/10.1007/978-3-642-70911-1\_20}

\bibitem{binney2008}
J.~{Binney}, S.~{Tremaine}, \emph{Galactic Dynamics}, 2nd edn.
\newblock Princeton Series in Astrophysics (Princeton University Press, 2008).
\newblock
  \urlprefix\url{http://gen.lib.rus.ec/book/index.php?md5=cd0fd2e719d8966f78eee1f04eee540e}

\bibitem{Kingma2015}
D.P. Kingma, J.L. Ba, in \emph{3rd International Conference on Learning
  Representations, ICLR 2015 - Conference Track Proceedings} (International
  Conference on Learning Representations, ICLR, 2015)

\bibitem{Zeiler2012}
M.D. {Zeiler}, arXiv e-prints arXiv:1212.5701 (2012)

\bibitem{Kiefer1952}
J.~Kiefer, J.~Wolfowitz, Annals of Mathematical Statistics \textbf{23}(3), 462
  (1952).
\newblock \doi{10.1214/AOMS/1177729392}

\bibitem{Robbins1951}
H.~Robbins, S.~Monro, Annals of Mathematical Statistics \textbf{22}(3), 400
  (1951).
\newblock \doi{10.1214/AOMS/1177729586}

\bibitem{Graham2005}
A.W. Graham, S.P. Driver, Publications of the Astronomical Society of Australia
   (2005).
\newblock \doi{10.1071/AS05001}

\bibitem{Roy2018}
N.~Roy, N.R. Napolitano, F.~{La Barbera}, C.~Tortora, F.~Getman, et~al.,
  Monthly Notices of the Royal Astronomical Society  (2018).
\newblock \doi{10.1093/mnras/sty1917}

\bibitem{Breiman2001}
L.~Breiman, Machine Learning \textbf{45}(1), 5 (2001).
\newblock \doi{10.1023/A:1010933404324}

\bibitem{Alam2015}
S.~{Alam}, F.D. {Albareti}, C.~{Allende Prieto}, F.~{Anders}, S.F. {Anderson},
  et~al.
\newblock {The Eleventh and Twelfth Data Releases of the Sloan Digital Sky
  Survey: Final Data from SDSS-III} (2015).
\newblock \doi{10.1088/0067-0049/219/1/12}

\bibitem{scikit-learn}
F.~Pedregosa, G.~Varoquaux, A.~Gramfort, V.~Michel, B.~Thirion, et~al., Journal
  of Machine Learning Research \textbf{12}, 2825 (2011)

\bibitem{brescia2019}
M.~{Brescia}, M.~{Salvato}, S.~{Cavuoti}, T.T. {Ananna}, G.~{Riccio}, S.M.
  {LaMassa}, C.M. {Urry}, G.~{Longo}, MNRAS \textbf{489}(1), 663 (2019).
\newblock \doi{10.1093/mnras/stz2159}

\end{thebibliography}

\end{document}